\documentclass[epj]{svjour}
\usepackage{graphics}
\usepackage{epsfig,array}
\newcommand{\be}{\begin{eqnarray}}
\newcommand{\ee}{\end{eqnarray}}
\newcommand{\er}{$\pm$}

\begin{document}
\title{The study of the negative pion production
in neutron-proton collisions at beam momenta below 1.8 GeV/c}
\author{
     V.V. Sarantsev   \inst{1}
\and K.N. Ermakov     \inst{1}
\and L.M. Kochenda    \inst{1}
\and V.I. Medvedev    \inst{1}
\and V.A. Nikonov     \inst{1,2}
\and O.V. Rogachevsky \inst{1}
\and \\
     A.V. Sarantsev   \inst{1,2}
\and S.G. Sherman     \inst{1}
\and V.A. Trofimov     \inst{1}
\and A.A. Vasiliev    \inst{1}
}
\institute{
Petersburg Nuclear Physics Institute, Gatchina 188300, Russia
\and  Helmholtz--Institut f\"ur Strahlen-- und Kernphysik,
Universit\"at Bonn, Germany
}

\date{}
\abstract {A detailed investigation of the reaction {\it np
$\rightarrow$ pp$\pi^{-}$} has been carried out using the data
obtained with the continuous neutron beam produced by charge
exchange scattering of protons off a deuterium target. A partial
wave event-by-event based maximum likelihood analysis was applied to
determine contributions of different partial waves to the pion
production process. The combined analysis of the $np \rightarrow pp
\pi^{-}$ and  {\it pp $\rightarrow$ pp$\pi^{0}$} data measured in
the same energy region allows us to determine the contribution of
isoscalar partial waves ({\it I=0}) in the momentum range from 1.1
up to 1.8 GeV/c. The decay of isoscalar partial waves into
$(^1S_0)_{pp}\pi$ channel provides a good tool for a determination of
the $pp$ S-wave scalar scattering length in the final state which
was found to be $a_{pp}=-7.5\pm 0.3$ fm.}

\authorrunning{V.V. Sarantsev et al.}
\titlerunning{The study of the negative pion production...}
\maketitle
\section{Introduction}
\label{intro}
The single pion production in $NN$ collisions is the main inelastic
process at energies below 1 GeV. Despite the fact that a lot of
experiments have been performed in this region, many questions on
this process are still far from being satisfactory answered. One of
them is the question about contributions of isoscalar ({\it I=0})
partial waves to the inelastic neutron-proton collision. The
neutron-proton scattering amplitude contains both isoscalar and
isovector ({\it I=0,1}) parts, and while the isovector part is
rather well known, even the magnitude of the total isoscalar
cross-section is badly determined. Usually, this cross section is
extracted from the difference of the total cross-sections of the
pion production reactions: {\it np $\rightarrow$ pp$\pi^{-}$} and
{\it pp $\rightarrow$ pp$\pi^{o}$}. However the numbers obtained in
different experiments are fairly scattered and do not give the whole
picture for the behavior of the isoscalar cross-section in the
energy region below 1 GeV. The only conclusion which can be made
from these values is that the contribution of the isoscalar cross
section to the $np\to pp\pi^-$ reaction is  smaller by about one
order of magnitude than the contribution of the isovector one.

It should be noted that the experimental data on the {\it np
$\rightarrow$ pp$\pi^{-}$} reaction below the energy of 1 Gev are
much poorer than those for {\it pp} collisions. The main reason for
this situation is due to the difficulty to create a pure
monoenergetic neutron beam. Therefore most previous experiments
used either a continuous neutron beam in a fairly wide energy range
\cite{1} or investigated the {\it pn}-interaction with a proton beam and a
deuteron target as a neutron source \cite{2,3,4}. Of course, it is
worth to mention the works where the energies of neutrons were defined
using the time-of-flight technique \cite{5,6}.

In the present study we use the continuous neutron beam produced by
the $pd$ interactions on the deuterium target. For a determination
of the contributions of the isoscalar cross-section we performed a
partial wave analysis of the data in whole region of incident
neutron momenta combined with the  $pp\to pp\pi^0$
data. In such an approach, the contribution of the isoscalar channel can
be determined, for example, from the asymmetry of the pion
distribution in the c.m.s. of the reaction, which is defined by the
interference of isoscalar and isovector amplitudes. Thus the
small isoscalar amplitudes can be defined with good accuracy.

\section{Experimental set up and data selection}

 The $np\to pp\pi^-$ data were taken at the PNPI synchrocyclotron with
the 35 cm hydrogen bubble chamber disposed in the 1.48 T magnetic
field. The neutrons were produced from collision of 1 GeV energy proton beam
with thin-walled liquid deuterium target.
The charged particles are swept away from the created neutron beam
by a system of the additional magnets and lenses situated behind the
deuteron target. The neutron beam passed through a collimator with
size 100x2x1 cm$^{3}$ and then the distance of 5m to the bubble
chamber. The produced neutrons had a momentum spread in a range
around $1\div2$ GeV/c. The detailed description of the neutron beam
and its energy distribution can be found in \cite{7}.

A total of $10^5$ stereofilms were obtained with
this$\hspace{0.44mm}$beam. The films were scanned twice with a goal to
select events with two positive curvature tracks and one negative
track. Due to small density of tracks on frames the efficiency of such
selection was near to 100$\%$. As the result, 10835 three prong events
were found. Those events can belong either to the single pion
production reaction
\begin{equation}
         np \rightarrow pp \pi^{-}
\end{equation}
or to the double pion production reactions
\be
           &np&\rightarrow np \pi^{+}\pi^{-},\\
           &np& \rightarrow d \pi^{+} \pi^{-}\,.
\ee
To separate the first process from the other two, tracks of events
disposed in the fiducial volume of the chamber were measured,
geometrically reconstructed and the momenta of assumed particles
were determined in according with their masses. The identification
of the events was performed by a kinematic fit demanding that the
confidence level of accepted events is better then 1\%. The
direction of the incident neutron beam is known from the collimating
system which provides the direction with the angular spread of
$0.3^{o}$ \cite{7}. But nothing is known about a value of the
neutron momentum which had to be defined by a 3C-constrained fit.
Three prong events were constrained by the kinematical fits for all
three mentioned above reactions (1)-(3). A visual scan of a bubble
density on the tracks was carried out for events fitted by several
hypotheses in order to distinguish between pion, proton and
deuteron. The angular resolution of the direction of the incident
neutron beam both in azimuthal and in polar angles was first
estimated to the $1^{o}$ which turned out to be more than three
times larger than that obtained after the fit. After repeated fits
with corrected errors we selected 8251 events assigned to reaction
(1). In the present experiment we did not monitor the neutron beam
and therefore the momentum distribution is given not in milibarns
but in the number of events. To determine the absolute isoscalar
cross section we included  in our partial wave analysis the {\it pp
$\rightarrow$ pp$\pi^{o}$} data for which the total cross section
was measured in our previous experiments.
\begin{figure}
\centerline{\epsfig{file=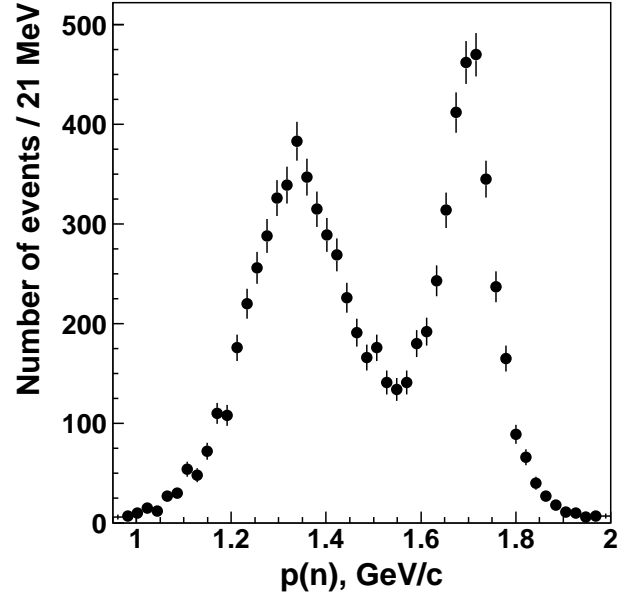,width=0.46\textwidth}}
\caption{ The neutron beam distribution for events of the reaction
{\it np $\rightarrow$ pp$\pi^{-}$}.}
\label{fig:1}       
\end{figure}

Fig.~\ref{fig:1} shows the neutron beam distribution for selected
$np \to pp\pi^-$ events. The neutron momentum distribution shows a
peak approximately at the energy of the original
protons incident on the deuterium target. The second peak with less
energetic neutrons is created in inelastic processes.

\begin{figure}
\centerline{
\epsfig{file=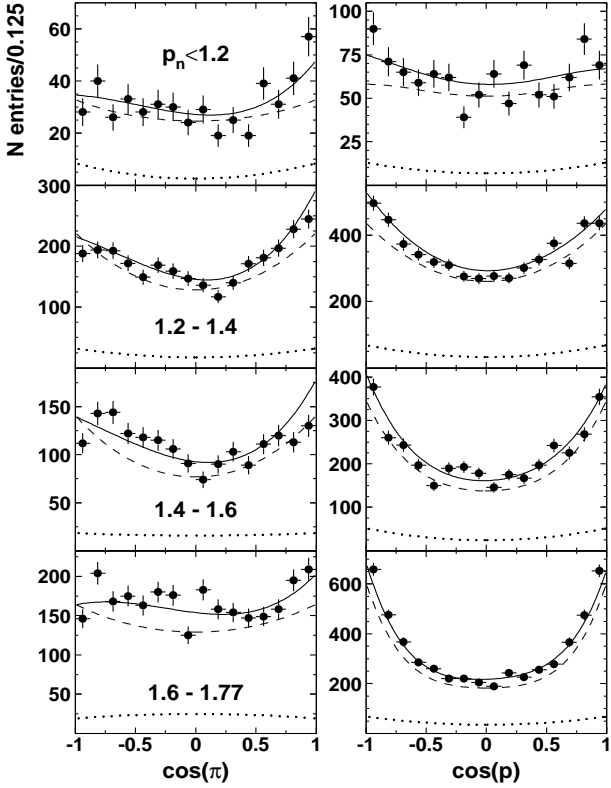,width=0.46\textwidth,height=0.45\textheight}~~
} \caption{The angular distribution of pion (left-hand column) and
final proton (right-hand column) in the c.m.s. of the $np\to
pp\pi^-$ reaction for 4 intervals of neutron momentum (in GeV/c).
The full curves are the result of the partial wave analysis, the
dashed curves are contributions from isovector amplitudes and dotted
curves are contributions from isoscalar amplitudes.}
\label{fig:2}       
\end{figure}

Angular distributions of pions and protons in the c.m.s. of the
reaction for chosen intervals of the initial neutron mo\-menta
are shown in Fig.~\ref{fig:2}. The proton angular distri\-bu\-ti\-ons
are rather symmetrical due to permutation of protons, while
the pion angular
distributions show a clear asymmetry (especially at low momenta of
incident neutrons). While isovector amplitudes are symmetrical with
respect to the direction of the produced pion, the isoscalar
am\-plitudes are antisymmetrical ones. Therefore, the asymmetry in the
differential cross section is defined by the interference of these
amplitudes providing a sensitive tool for a determination of the weak
isoscalar partial waves.

\begin{figure}
\centerline{
\epsfig{file=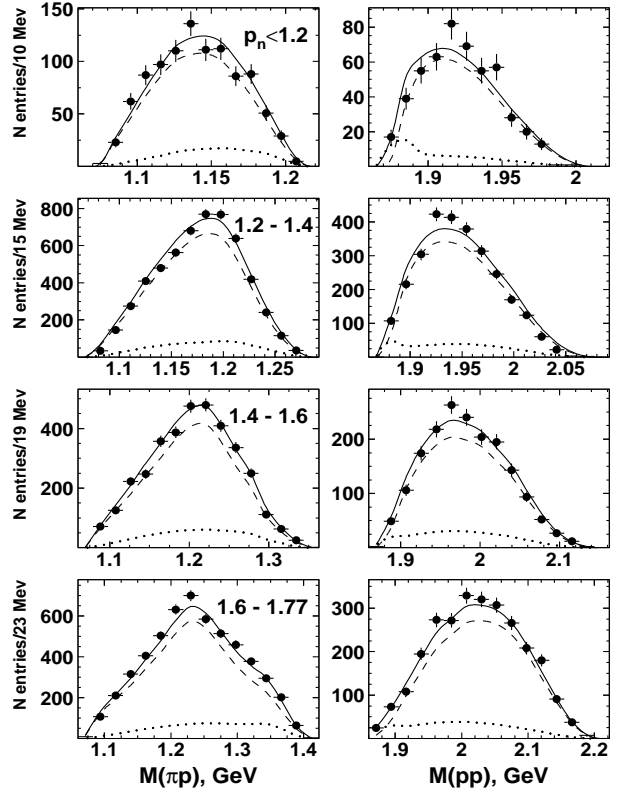,width=0.46\textwidth,height=0.45\textheight}~~
} \caption{The $\pi p$ and $pp$ invariant mass distributions of the
final particles for 4 intervals of neutron momentum (in GeV/c). The
full curves are the result of the partial wave analysis, the dashed
curves are contributions from isovector amplitudes and dotted curves
contributions from isoscalar amplitudes.}
\label{fig:3}       
\end{figure}

The two body invariant mass distributions of final particles are
shown in Fig.~\ref{fig:3}. The $\pi p$ invariant mass at high
energies of the neutrons has a peak defined by the produc\-tion of
$\Delta(1232)$ state and indeed, as the partial wave ana\-ly\-sis
has shown, the $\Delta(1232)p$ channel is one of dominant final
states in this reaction. The $pp$ invariant mass distributions are
smooth and do not reveal any peculiarities.

\section{Parameterization of partial waves}

For the production of three particles with the 4-momenta $q_i$ from
two particles colliding with 4-momenta $k_1$ and $k_2$, the cross
section  is given by:
\be
d\sigma=\frac{(2\pi)^4|A|^2}{4|\vec k|\sqrt{s}}\,
d\Phi_3(P,q_1,q_2,q_3)\;, \qquad P\!=\!k_1\!+\!k_2\;,
\ee
where $A$ is the reaction amplitude, $\vec k$ is the 3-momentum of
the initial particle calculated in the c.m.s system  of the
reaction, and $s=P^2=(k_1+k_2)^2$. The invariant three particle
phase space is given by
\be
d\Phi_m(P,q_1\ldots q_m)\! =\! \delta^4(P- \sum\limits_{i=1}^3
q_i)\! \prod\limits_{i=1}^3 \! \frac{d^3q_i}{(2\pi)^3 2q_{0i}}\;.
\label{phase}
\ee
The total amplitude can be written as sum of the partial wave
amplitudes:
\be
A=\sum\limits_\alpha A^\alpha_{tr}(s) Q^{in}_{\mu_1\ldots\mu_J}(S
LJ)A_{2b}(i,S_2L_2 J_2)(s_i)\times
\nonumber\\
Q^{fin}_{\mu_1\ldots\mu_J}(i,S_2L_2J_2S'L'J)\ .
\ee
Here $S,L,J$ are spin, orbital momentum and total angular momentum
of the $np$ system, $S_2,L_2,J_2$ are spin, orbital momentum and
total angular momentum of a two-particle system in the final state and $S',L'$
are spin and orbital momentum between the two particle system and a
third particle with momentum $q_i$. The invariant mass of the two
body system can be calculated as $s_i=(P-q_i)^2$. The multiindex
$\alpha$ denotes all possible combinations of the
$S,L,J,S_2,L_2,J_2,S',L'$ and $i$,  $A^\alpha_{tr}(s)$ is the
transition amplitude and $A_{2b}(i,S_2L_2 J_2)(s_i)$ describes
rescattering processes in the final two particle channel (e.g.
production of $\Delta(1232)$). In this spin-orbital momentum
decomposition we follow the formalism given in \cite{8,9,10}. The
exact form of the operators for initial states
$Q^{in}_{\mu_1\ldots\mu_J}(S LJ)$ and final states
$Q^{fin}_{\mu_1\ldots\mu_J}(i,S_2L_2J_2S'L'J)$ can be found in
\cite{10}.

Following this decomposition we use the usual spectroscopic notation
$^{2S+1}L_{J}$ for description of an initial state, the system of
two final particles and  the system "spectator  and two-final
particle state". For the initial $np$ system, states with total
momenta $J \leq 2$ and angular momenta between two nucleons of
$L=0,1,2,3$ were taken into account. For the final three particle
system we restricted the fit by angular momenta $L_2=0,1,2$ and
$L'=0,1,2$.

Due to the nonresonant nature of the $np$ system in the energy
region investigated here, there is no factorization between initial
and final vertices and the transition amplitude can depend on all
quantum numbers which characterize a partial wave (index $\alpha$).
Moreover, due to contribution of the triangle singularities, the
production parameters can be complex numbers. The best description
was obtained with the parameterization
\begin{equation}
A^\alpha_{tr}(s)=\frac{a^\alpha_{1}+a^\alpha_{3}\sqrt{s}}
{s-a^\alpha_4}\, e^{ia^\alpha_{2}},
\label{trans}
\end{equation}
where $a^\alpha_i$ are real parameters. The $a^\alpha_4$ parameter
corresponds to a pole situated in the region of left-hand side
singularities of the partial wave amplitudes. It is introduced to
suppress the growing of the amplitudes at the regions of large $s$.

We also used other, more complicated parameterizations of the
transition amplitude. However, either we obtained a worse
description of the data or a similar description with larger number
of parameters in the fit. In the latter case we included those
results to determine the final systematical errors in the isoscalar
contribution to the cross section.

\begin{figure}
\centerline{ \epsfig{file=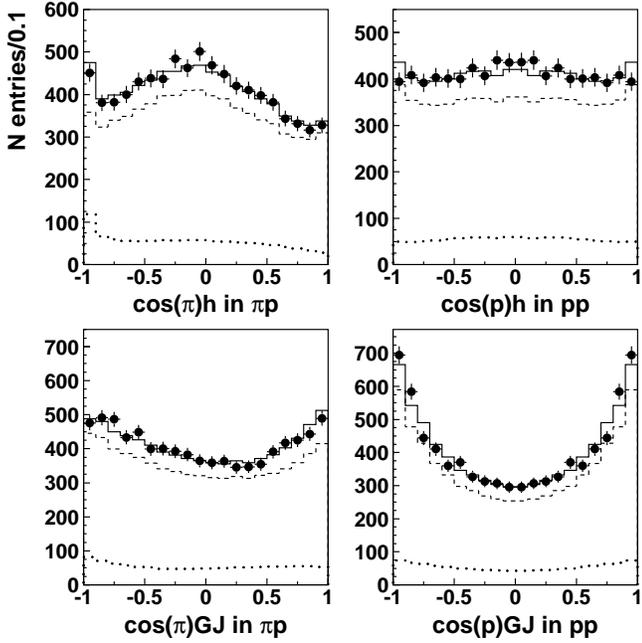,width=0.48\textwidth} }
\caption{The angular distribution of the final pion and proton in
the helicity and Godfrey-Jackson system for the total interval of
neutron momentum. The full histograms are the result of the partial
wave analysis, the dashed histograms are the contributions from
isovector amplitudes and dotted histograms are contributions from
isoscalar amplitudes.}
\label{fig:4}       
\end{figure}

For the $\pi p$ system in the intermediate state we introduce
two resonances, $\Delta(1232)P_{33}$ and Roper $N(1440)P_{11}$. For
the $\Delta(1232)$, we use a relativistic Breit-Wigner formula with
mass and width fixed from the PDG values. The Roper state was
parameterized in agreement with Breit-Wigner couplings found in the
analysis \cite{our_roper}. Let us note that
the present analysis is not
sensitive to the exact parametrization of the Roper resonance: only the low energy
tail of this state can influence the data.

For the description of the final $pp$ interaction we use a
modified scattering length approximation formula:
\begin{equation}
A_{2b}^{\beta}(s_i)= \frac{\sqrt{s_i}}{1-\frac 12 r^\beta
q^{2}a^\beta_{pp}+ iqa^\beta_{pp}q^{2L}/F(q,r^\beta,L)},
\label{a_2b}
\end{equation}
where multiindex $\beta$ denotes possible combinations of a
kinematical channel $i$ and quantum numbers  $S_2$, $L_2$ and $J_2$,
$a_{pp}^\beta$ is a $pp$-scattering length and $r^\beta$ is the
effective range of the $pp$ system. The $F(q,r,L)$ is the Blatt-Weiss\-kopf
form factor (it is equal to 1 for $L=0$ and the explicit form for
other partial waves can be found, for example in \cite{8}) and $q$
is a relative momentum in the $pp$-system:
\begin{equation}
q=\frac{\sqrt{s_i-4m_p^2}}{2}
\end{equation}

For the S-wave partial wave this formula corresponds exactly to the
scattering length approximation suggested in \cite{watson,migdal}.
To check a possible additional energy dependence we also introduced
a more complicated parameterization of the numerator in Eq. (\ref{a_2b})
by substituting:
\be
\sqrt{s_i}\to \sqrt{s_i}+b^\beta\,s_i
\ee
where $b^\beta$ are fit parameters. However we found that this
freedom does not improve the description but leads to large
correlations between these parameters and parameters of the
transition amplitude $A^\alpha_{tr}(s)$.

\section{The results and discussion}

We minimized the log-likelihood value fitting simultaneously the
present data taken in the whole range of the neutron momenta
and data obtained earlier on $pp\to pp\pi^0$ \cite{14,15}
measured
at nine energies covering the same energy interval as the $np$ data.
To fix the low energy region one set of high statistics
data taken by the T\"ubingen group \cite{tubingen} on $pp\to
pp\pi^0$ was also introduced to the fit. The $pp$
collision events are produced solely from the isovector channel
($I=1$) and therefore the contribution of isovector states to the
$np\to pp\pi^-$ reaction (which differs only by a common
Clebsch-Gordan coefficient) is strongly constrained. This
appreciably increases the accuracy for the extraction of the
isoscalar contribution from the data.

\begin{figure}
\centerline{\epsfig{file=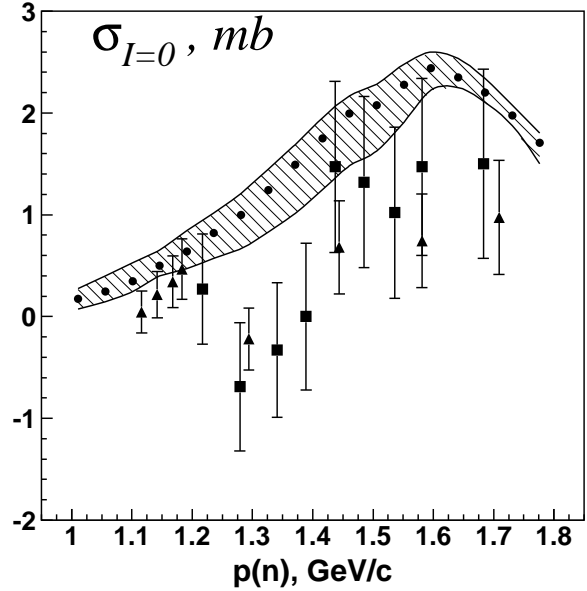,width=0.45\textwidth} }
\caption{The
isoscalar total cross section. The band represents the systematical
errors obtained from the large number of fits. The experimental
points are taken from \cite{15} (squares) and \cite{17} (triangles).}
\label{fig_tcs}       
\end{figure}

The results of the partial wave analysis are shown in
Figs.~\ref{fig:2}-\ref{fig:3} by solid curves. Contributions of
the isovector channel are given by dashed curves and contributions
from isoscalar channel as dotted curves. It is seen that the fit
reproduces well the asymmetry of the pion angular distribution in
the overall c.m.s. (see Fig.~\ref{fig:2}). Moreover,
the angular distribution of isoscalar channel changes notably with
energy demonstrating contributions from different partial waves.

The largest contributions from isovector initial states to the pion
production process stem from the $^3P_2$, $^3P_1$ waves (see
Table~\ref{table1}). At large neutron momenta the $^3F_2$
contribution is also significant. The $^1S_0$ wave gives the
ne\-gligible contribution.
The $^3P_2$ state decays almost
equal\-ly into the $\Delta(1232)p$, $N(1440)P_{11}p$ and
$(^3P_2)_{pp}\pi$ states. The $^3P_1$ initial state decays$\hspace{2mm}$
also$\hspace{2mm}$ almost equally into $\Delta(1232)p$ and
$(^3P_2)_{pp}\pi$, while $^3F_2$ decays dominantly into the
$\Delta(1232)p$ and $N(1440)P_{11}p$ final states. The dominant
isoscalar sta\-te $^1P_1$ decays dominantly into $(^3P_1)_{pp}\pi$ and
$N(1440)P_{11}p$, while $^3S_1$ and $^3D_1$ states decay almost
totally into the $(^1S_0)_{pp}\pi$ channel.

\begin{table}[pt]
\begin{center}
\caption{\label{table1} Largest contributions of initial partial waves
($^{2S+1}L_J$) to the reaction $np\to pp\pi^-$. The central values
are given for the best fit and errors are extracted from the large
number of different solutions. \vspace{2mm}}
\begin{tabular}{cc|cc}
\hline
\hline
\multicolumn{2}{c}{$I=1$} &\multicolumn{2}{c}{$I=0$}\\
\hline
$^3P_0$ & 3.7\er 1.3\,\%  & $^3S_1$ & 0.9\er 0.2\,\% \\
$^3P_1$ & 21.6\er 3.0\,\%  & $^1P_1$ & 10.9\er 1.2\,\%  \\
$^3P_2$ & 46.1\er 5.0\,\%  & $^3D_1$ & 1.8\er 0.3\,\%  \\
$^1D_2$ & 4.7\er 1.5\,\%  & ~ & ~ \\
$^3F_2$ & 10.3\er 2.5\,\%  & ~ & ~ \\
\hline
\hline
\end{tabular}
\end{center}
\end{table}

The quality of the partial wave analysis is also demonstrated in
Fig.~\ref{fig:4}. Here, angular distributions of the pion and final
proton are compared to the fit in the helicity and Godfrey-Jackson
frames for the total momentum range of the incident neutron. The
helicity frame is a rest frame of the two final particles with the
angle calculated between one of constituent particles and a
spectator particle. This frame is most suitable for the
investigation of processes when two initial particles form a system
(e.g. resonance) which decays into a final two-body system and a
spectator. The Godfrey-Jackson frame is defined as rest frame of two
final particles, however here the angle is calculated between one of
constituent particles and the beam. This system is mostly suitable
for the study of two particle systems in the final state produced by
$t$-channel exchange. We would like to mention that the description
of these distributions at the same momentum intervals as given in
Figs.~\ref{fig:2},\ref{fig:3} has the same quality as a total
distribution.

\begin{figure}
\centerline{ \epsfig{file=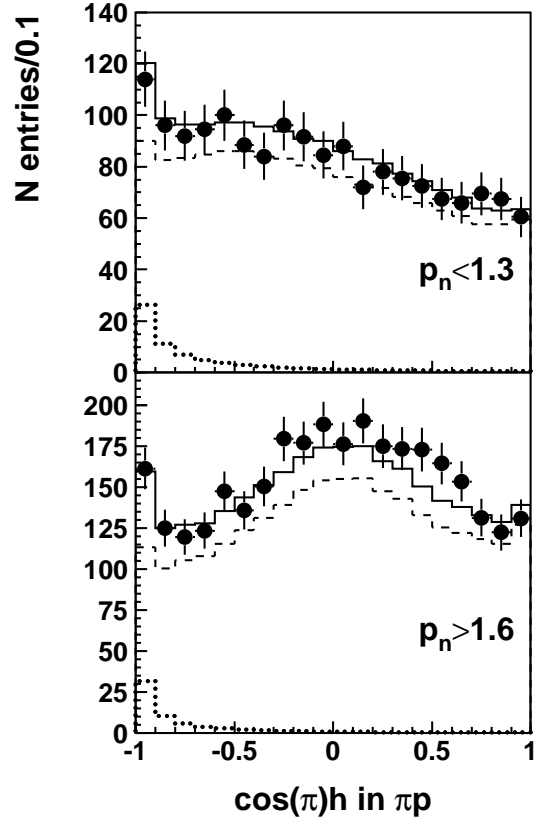,width=0.40\textwidth}~~ }
\caption{The angular distribution of the final pion in the helicity
system for two intervals of incident neutron momentum (in GeV/c).
The full histograms (close to the data points) are the result of the
partial wave analysis, the dashed histograms show the contributions
from isovector channel and dotted histograms the contributions from
isoscalar channel decaying into the $(^1S_0)_{pp}\pi$ final state.}
\label{fig_hel}       
\end{figure}

The partial wave analysis defines the relative contribution of the
isovector and isoscalar channels to the total cross section. The
isovector total cross section was measured experimentally \cite{14,15}
which provides the possibility to define the energy dependence of
the total cross section for the $np\to pp\pi^-$ reaction.

A comparison of the isoscalar cross section found in the partial
wave analysis with results from other experiments extracted as:
\begin{equation}
\sigma(I=0)=3[2\sigma(np \rightarrow pp\pi^-)-\sigma(pp \rightarrow
pp\pi^0)] \,,
\end{equation}
is shown in Fig.~\ref{fig_tcs}. The points inside the band
correspond to our best solution. The band itself shows the
systematical errors obtained from a large set of solutions with
different parameterizations of partial wave amplitudes. It is
seen that our result is systematically exceeds the previous
calculations.

The isoscalar cross section increases smoothly from 1.1 GeV/c up to
1.6 GeV/c where it reaches a value of 2.5 mb and then it drops
sharply down to 1.5 mb. It is possible that this drop is connected
with the opening of double pion production channels where isoscalar
$np$ states could decay, for example, into two $\Delta$ states.

The isoscalar initial channel provides a very good tool for a
determination of the scattering length of the final $pp$ system in
the pion production reactions. In the initial fits we fixed
scattering length and effective range of the $^1S_0$ final $pp$
state to $a_{pp}=-7.83$ fm and $r=2.8$ fm \cite{18}. For
other partial waves the effective ranges were also fixed to $r=2.8$
fm while parameters $a^\beta_{pp}$ were fitted freely.

An appreciable fraction of the isoscalar channel decays into
pion and the $pp$ system  in the final $^1S_0$ state. The $^1S_0$ state
produces a sharp peak in the pion backward angle in the helicity
frame. This angle corresponds to the lowest $pp$ invariant
mass and due to kinematical conditions
(suppression of other partial waves), this peak is well observed in
the data. The pion angular distributions in the helicity frame for
two intervals of the incident neutron momentum is shown in
Fig.~\ref{fig_hel}. It is seen that the isoscalar channel produces a
dominant effect at small and large momenta of incident neutron. If
the effective range is fixed at 2.8 fm, we found the scattering
length to be equal to $a_{pp}=-7.5\pm 0.3$ fm.


\section{Conclusions}

We performed a detailed study of the $np \rightarrow pp \pi^{-}$
reaction. The data were analyzed in a maximum likelihood partial
wave analysis. The partial wave analysis provides a solid
determination of weak isoscalar amplitudes from their interference
with strong isovector amplitudes. The dominant decay of the initial
isoscalar states $^3S_1$ and $^3D_1$ into the $(^1S_0)_{pp}\pi$
final state allows us to obtain a good determination of the $pp$
scattering length in the final state of pion production reactions.

 \section{Acknowledgements}

   We would like to express our gratitude to the bubble
chamber staff as well as to the laboratory assistants, which toiled
at the film scanning and measuring. We are grateful E. Klempt for
useful discussions and reading the paper. Part of the work was
supported by a FFE grant of the Re\-se\-arch Center J\"ulich and by
the Deutsche Forschungs\-ge\-mein\-schaft within the
Sonderforschungs\-be\-reich SFB/TR16.

\end{document}